\documentclass[final,5p,times,twocolumn]{elsarticle}

\usepackage{graphicx}

\usepackage{amssymb}
\usepackage{amsmath}
\usepackage{hyperref}
\hypersetup{
	colorlinks = true,
	linkcolor = Blue,
	citecolor = Blue,
	urlcolor  = Blue
}

\journal{Computer Physics Communications}

\begin{document}

\begin{frontmatter}



\title{MuFinder: A program to determine and analyse muon stopping sites}


\author[durham]{B. M. Huddart\corref{author}}
\author[durham]{A. Hern\'{a}ndez-Meli\'{a}n}
\author[durham]{T. J. Hicken\fnref{rhul}}
\author[durham,ijs,llubljana]{M. Gomil\v{s}ek}
\author[durham]{Z. Hawkhead}
\author[durham]{S. J. Clark}
\author[isis]{F. L. Pratt}
\author[durham]{T. Lancaster}

\address[durham]{Centre for Materials Physics, Durham University, Durham DH1 3LE, United Kingdom}

\address[ijs]{Jo\v{z}ef Stefan Institute, Jamova c. 39, SI-1000 Llubljana, Slovenia}

\address[llubljana]{Faculty of Mathematics and Physics, University of Ljubljana, Jadranska u. 19, SI-1000 Llubljana, Slovenia}

\address[isis]{ISIS Pulsed Neutron and Muon Facility, STFC Rutherford Appleton Laboratory, Harwell Oxford, Didcot OX11 OQX, United Kingdom}

\cortext[author] {Corresponding author.\\\textit{E-mail address:} benjamin.m.huddart@durham.ac.uk}
\fntext[rhul] {\textit{Present address:} Department of Physics, Royal Holloway, University of London, Egham TW20 0EX, United Kingdom}

\begin{abstract}
Significant progress has recently been made in calculating muon stopping sites using density functional theory. The technique aims to address two of the most common criticisms of the muon-spin spectroscopy ($\mu^+$SR) technique, namely, where in the sample does the muon stop, and what is its effect on its local environment. We have designed and developed a program called MuFinder that enables users to carry out these calculations through a simple graphical user interface (GUI). The procedure for calculating muon sites by generating initial muon positions, relaxing the structures, and then clustering and analysing the resulting candidate sites, can be done entirely within the GUI. The local magnetic field at the muon site can also be computed, allowing the connection between the muon sites obtained and experiment to be made. MuFinder will make these computations significantly more accessible to non-experts and help to establish muon site calculations as a routine part of $\mu^+$SR experiments. 
\end{abstract}


\end{frontmatter}



{\bf PROGRAM SUMMARY}

\begin{small}
\noindent
{\em Program Title:} MuFinder                                         \\
{\em CPC Library link to program files:} (to be added by Technical Editor) \\
{\em Developer's repository link:} https://gitlab.com/BenHuddart/mufinder \\
{\em Code Ocean capsule:} (to be added by Technical Editor)\\
{\em Licensing provisions:} GPLv3 \\
{\em Programming language:}  Python                                 \\
{\em Nature of problem:}\\
 To automate the process of calculating muon stopping sites using density functional theory, thereby making these calculations accessible to non-experts.\\
{\em Solution method:}\\
A Python-based graphical user interface (GUI) through which users can calculate muon stopping sites using the structural relaxation method. The program makes use of newly-developed algorithms for generating candidate initial muon positions and for clustering muon positions obtained from the structural relaxations. Analysis of the muon sites, including calculation of the local dipolar magnetic field, is also possible within the GUI.
   \\
%
\end{small}

\section{Introduction}
Muon-spin spectroscopy ($\mu^+$SR) is an experimental technique in which spin-polarised muons are implanted in a sample, with the time evolution of their polarisation providing information about the local magnetic fields present \cite{muon_book}. Two of the most fundamental limitations of this technique are the lack of knowledge of the muon stopping site, and the uncertainty surrounding the degree to which the muon distorts its local environment.
In some cases it has been possible to determine the muon stopping site experimentally: through the angular dependence of the muon frequency shift in an applied field \cite{PhysRevB.30.186,Amato1997,PhysRevB.89.184425}, level-crossing resonances \cite{PhysRevLett.60.224,Brewer1991} or from the entanglement between the muon's spin and the spins of a small number of surrounding nuclei \cite{PhysRevB.33.7813,PhysRevLett.99.267601}.  The subset of systems for which each of these approaches are applicable is limited and they therefore do not represent general methods for muon site assignment.  However, there has recently been significant progress in calculating muon stopping sites using {\it ab initio} methods, particularly with density functional theory (DFT), a procedure which has been come to be known as DFT+$\mu$ (see Ref.~\cite{hideseek} for a review). Knowledge of the muon stopping site can allow one to constrain the sizes \cite{PhysRevB.91.144417} and/or directions \cite{PhysRevB.100.094401} of ordered moments, or to fit experimental data to models that depend quantitatively on the local environment of the muon \cite{PhysRevB.87.115148,PhysRevLett.125.087201}.

There are two distinct approaches that have been used to determine muon stopping sites. In the Unperturbed Electrostatic Potential (UEP) method, the electrostatic potential of the host crystal is calculated using DFT.  For a positively charged defect such as $\mu^+$, the minima of the electrostatic potential are candidate stopping sites. This method has been found to give a good approximation for the muon stopping site in metallic systems \cite{PhysRevB.80.094524,De_Renzi_2012,Lamura_2013}, where screening of the $\mu^+$ charge prevents strong bonding.  However, for covalent or ionic systems, such as insulating fluorides \cite{moeller,PhysRevB.87.115148}, it is found the stable muon sites do not generally coincide with the minima of the electrostatic potential as a result of the strong muon--lattice interactions in these systems \cite{hideseek}. 
Moreover, the UEP method cannot be used to determine stopping sites for muonium (the bound state of a muon and an electron) as, being electrically neutral, there is no reason why it should necessarily localise in an electrostatic minimum.  An alternative approach based on structural relaxations provides a more robust method of determining muon stopping sites.  Here, the muon (modelled as a light proton) is placed in randomly-chosen low-symmetry sites in the structure and all of ions are then allowed to relax.  This is more computationally expensive than the UEP method, as each initial muon position requires a geometry optimisation calculation.  The computational cost is increased further by the fact that this approach often requires the use of supercells in order to minimise the interaction of the muon with its periodic images. In a $\mu^+$SR experiment muons are implanted in the ultradilute limit, so muon--muon interactions never take place, and hence our simulations must reflect this.
A strength of the structural relaxation approach is that it allows the muon-induced distortions of the host crystal to be evaluated, with these potentially having a significant effect on the response of the muon to the system under study.  A particularly striking case is that of Pr-based pyrochlores, where the anisotropic distortion field induced by the muon splits the crystal field levels of Pr$^{3+}$, resulting in a muon response that is dominated by the distortion it induces, rather than the intrinsic properties of the sample  \cite{PhysRevLett.114.017602}. On the other hand, muon-induced distortions in the spin ladder compound (Hpip)$_2$CuBr$_4$, determined using DFT,  are thought to be responsible for the sensitivity of the muon to the underlying magnetic state of the system \cite{spinladder}. Knowledge of the muon stopping site (obtained from either approach) makes it possible for $\mu^+$SR measurements to provide estimates for magnetic moment sizes or to compare different candidate magnetic structures, with a notable success being the determination of the helical \cite{PhysRevB.93.144419} and conical \cite{PhysRevB.95.180403} phases of MnSi.

In addition to many convincing applications of DFT+$\mu$, research has been done into making these techniques more accessible to experimentalists, who might lack proficiency in carrying out electronic structure calculations. This includes investigating the use of lower-level approximations to DFT, such as Density Functional Tight Binding (DFTB), to reduce the computational expense \cite{sturniolo} and the development of scripts and utilities to facilitate these calculations, such as those Muon Spectroscopy Computational Project software suite \cite{sturniolo}. However, work still remains to lower the barrier to entry for being able to perform these calculations and hence establish them as a routine part of carrying out a muon-spin spectroscopy experiment.
Here we introduce MuFinder, a program that facilitates the process of calculating muon sites using the structural relaxation method by allowing these calculations to be run through a graphical user interface (GUI), which can also be used to calculated the dipolar magnetic field at the obtained sites. In section~\ref{design} we summarise the principles behind the design of the program. In section~\ref{algorithms} we introduce new algorithms for determining and analysing muon stopping sites that are implemented in MuFinder. In section~\ref{usage} we demonstrate the usage of the program through the example of calculating muon sites in CoF$_2$. Finally, in section~\ref{conclusion} we summarise the benefits of the program and suggest avenues for future development.

\section{Design principles}\label{design}

The continuing growth of computational methods in physics has also prompted the development of software packages to act as front ends for many of these codes, to make these approaches accessible to a wider range of researchers. Many of these packages take the form of Python libraries, as the Python programming language is increasingly popular, due to it running on all operating systems and having an easy to read syntax. For example, the Atomic Simulation Environment (ASE)~\cite{ASE}, which is written in Python, provides a powerful means to manipulate, analyse and visualise atomic structures, and provides the basis for more specialised packages. These include the Soprano library~\cite{soprano} for handling collections of structures and the MuESR \cite{muesr} library for calculating the local magnetic field at a muon site. The existence of these specialist packages, combined with many of the other useful libraries available made Python the ideal programming language for developing our program.

\begin{figure}[ht]
\centering
\includegraphics[width=\columnwidth]{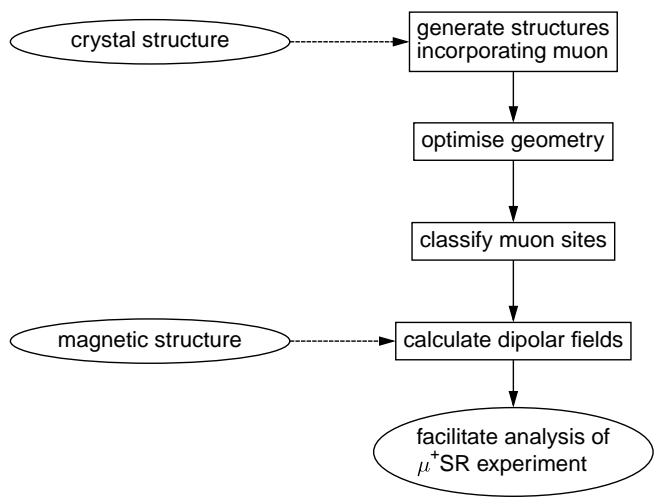}
\caption{The workflow for a muon site calculation using MuFinder.} \label{fig:fig1}
\end{figure}

\begin{figure*}[ht]
\centering
\includegraphics[width=\textwidth]{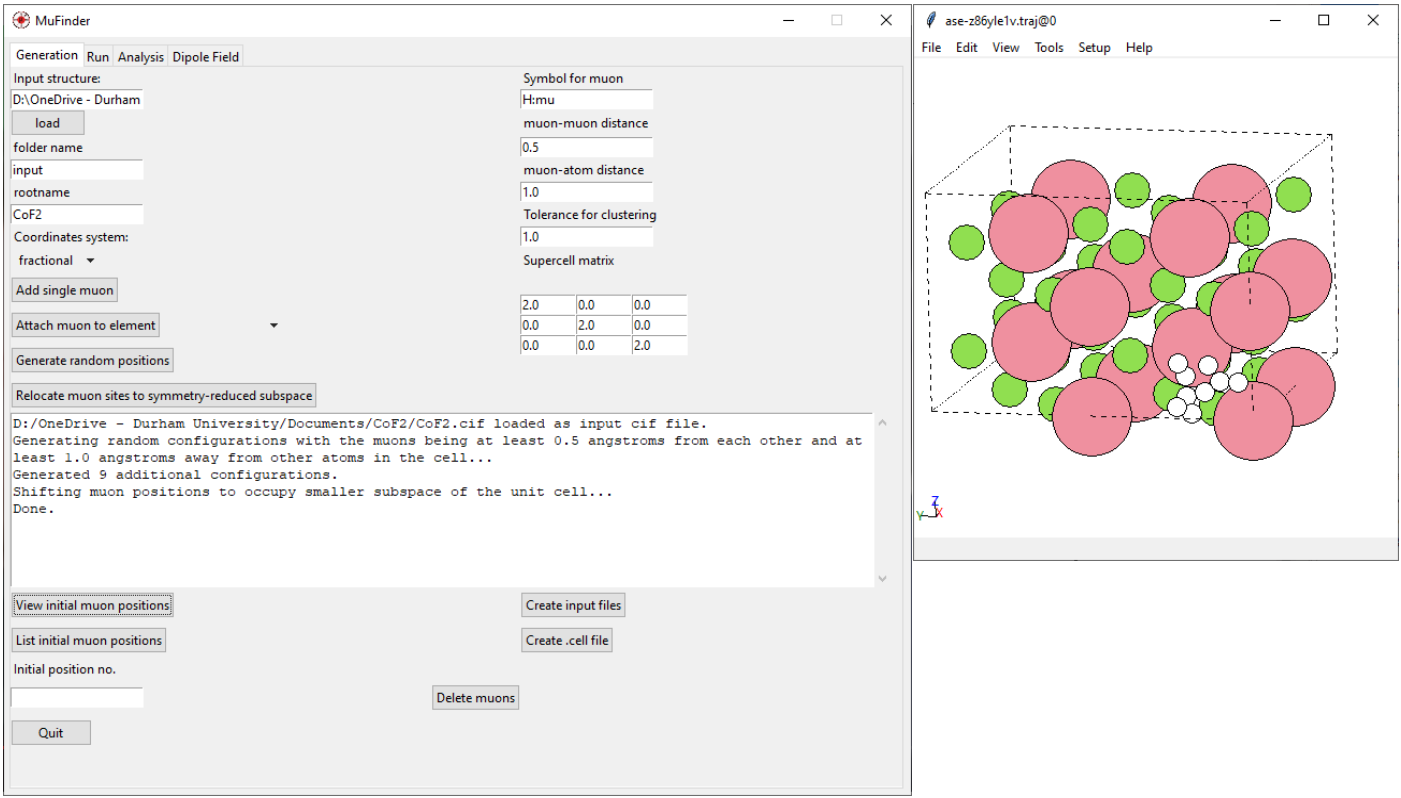}
\caption{A screenshot of the MuFinder program, demonstrating initial muon position generation. \textit{Left:} the main GUI window. \textit{Right:} initial muon positions in a $2\times2\times2$ supercell of CoF$_2$ visualised using ASE's GUI, which is built into the program.}
\label{fig:fig2}
\end{figure*}

MuFinder is a Python-based program that builds upon previously developed of methods for calculating muon stopping sites using DFT \cite{hideseek,liborio} by providing a GUI that aims to enable non-experts to carry out these calculations. It is designed to facilitate the following workflow (illustrated in Figure~\ref{fig:fig1}): firstly, given a crystal structure, candidate structures consisting of a muon embedded in the structure are generated; these structures are then relaxed, with MuFinder providing tools to run these calculations either locally or on a remote computing cluster; the relaxed structures are then clustered to identify distinct stopping sites; finally, the dipolar field at the muon stopping sites can be calculated if required. The MuFinder GUI [Figure~\ref{fig:fig2} (left)] comprises four tabs, with each tab corresponding to a step in this workflow. The user works their way along these tab from left to right in order to determine and then analyse the muon stopping sites. The contents and operation of each tab is explained in section~\ref{usage}.

Simplicity and ease of use was at the forefront in the design of MuFinder. The user needs only provide the structure of their system, which is done by supplying a Crystallographic Information File (CIF) \cite{CIF} file; no other input files are required. MuFinder provides an intuitive interface for configuration of the parameters for DFT calculations, with appropriate default values for this type of calculation. For more advanced users, it is also possible to manually edit the contents of the input files via a text window accessible through the GUI.

MuFinder uses the CASTEP \cite{CASTEP} electronic structure code to carry out geometry optimisation calculations. CASTEP is a fully-featured plane-wave basis set DFT code that is particularly user-friendly, owing to its very understandable syntax for input and output and sensible default values for input parameters. An existing CASTEP installation is required, and this can be installed either locally or on a remote machine. While only the CASTEP electronic structure code is currently supported, we note that ASE provides Python wrappers to a large number of codes including other plane-wave basis-set DFT codes such as \textsc{Quantum ESPRESSO}, Gaussian based electronic structure codes such as \textsc{Gaussian} and the DFT-based tight binding code DFTB+, the latter of which has been shown to be a less expensive method (using greater approximations) for determining muon site in organic molecular crystals \cite{sturniolo}. As the analysis tools in MuFinder work with ASE Atoms objects it will be straightforward to expand the program to work with a wider range of electronic structure codes.

\section{Algorithms} \label{algorithms}

\subsection{Initial position generation}\label{gen_alg}

In the structural relaxation approach to calculating muon stopping sites, the possibility of multiple local minima in the potential energy surface for the muon requires a number of initial muon positions to be sampled to successfully identify all of the distinct candidate sites.  Each of these initial muon positions must be relaxed and hence the computational cost increases linearly with the number of initial positions.  Thus it is important to be able to generate sets of initial positions that effectively sample the potential energy landscape while minimising the number of initial positions (and hence geometry optimisation calculations).

The algorithm used by MuFinder for generating initial muon positions is based on the one described in Ref. \cite{liborio} and is as follows:
\begin{enumerate}
\item Generate random positions within the conventional unit cell.
\item Accept each position if it and its symmetry equivalent positions are all:
\begin{enumerate}[(i)]
	\item at least $r_\textrm{muon}$ away from the other muon positions and
	\item at least $r_\textrm{atom}$ away from all of the atoms in the cell.
\end{enumerate}
\item Repeat until $n_\mathrm{rejected}$ new positions are rejected.
\end{enumerate}

The number of initial structure generated will depend on $r_\mathrm{muon}$ and $r_\mathrm{atom}$, with smaller values leading to a greater number of structures.  The choice of the muon--muon distance $r_\mathrm{muon}$ is dictated by the expected shape of the potential energy surface for the muon.  For a surface with a large number of minima, finer sampling of the unit cell will be required to successfully locate all of these minima.  The muon--atom distance $r_\mathrm{atom}$ should be chosen to exclude unphysical situations where the muon sits very close to an atom in the structure.  Values of $r_\mathrm{atom}$ can be chosen using physical intuition based on typical muon--ion distances, with the MuFinder default value $r_\mathrm{atom}=1.0$~\AA~being the typical $\mu^+-$O bond length.  A novel feature of the algorithm used by MuFinder to generate initial muon positions is the consideration of symmetry equivalent positions, which can significantly reduce the search space of the unit cell that needs to be sampled.  As the multiplicity of a generic position in the unit cell under the symmetry operations of crystal's space group is the same (provided it is not a high-symmetry point, which is extremely unlikely, i.e., occurs with zero probability, for a randomly generated position), the positions generated represent an unbiased sampling of the unit cell. This is important if one wishes to make inferences about the basin of attraction of each muon site from the number of initial structures that relax into this site. Increasing the number $n_\mathrm{rejected}$ of new positions that can be rejected before terminating the algorithm results in a set of muon positions that fills more of the space available in the crystal structure, at the expensive of generating a larger number of initial positions that will not be used in the calculations. A value $n_\mathrm{rejected}=30$ is currently used by MuFinder. Larger effective values of $n_\mathrm{rejected}$ can be achieved by running the algorithm multiple times and appending the additional initial positions generated each time. After generating a set of initial muon positions according to this algorithm, structures comprising the host structure and an implanted muon can be prepared for use as inputs for geometry optimisation calculations. 

\subsection{Clustering algorithm} \label{clustering}
\begin{figure}[t]
\centering
\includegraphics[width=\columnwidth]{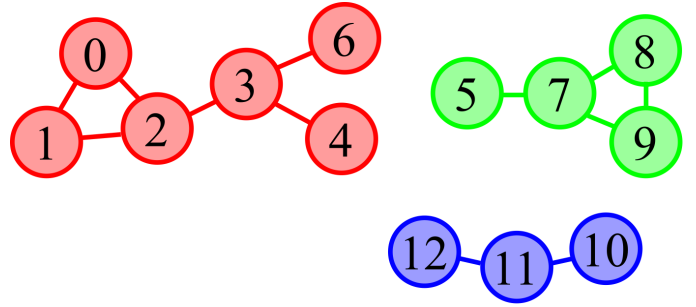}
\caption{A sample graph comprising a set of nodes (which could each represent a muon stopping site) connected by (undirected) edges.  The graph can be separated into three component subgraphs (coloured red, green and blue) in which any two nodes within the subgraph are connected to each other by paths, with no paths between nodes belonging to different subgraphs. These are the connected components of the graph.}
\label{fig:fig3}
\end{figure}

After carrying out the structural relaxations, one is left with a candidate muon site corresponding to each initial muon position. However, it is unlikely that every initial position will lead to a distinct stopping site in the final structure. A method for clustering the resulting sites in order to identify a smaller set of distinct muon sites is therefore required. The clustering algorithm used in MuFinder uses ideas from graph theory. To generate a graph for a set of relaxed muon positions $\mathbf{r}_i$, we first construct a distance matrix \textbf{D}, where the component $D_{ij}$ is the minimum distance between muon sites $i$ and $j$, taking into account symmetry equivalent positions, or, more formally,
\begin{equation}
D_{ij} = \min_T |\mathbf{r}_i - T \mathbf{r}_j| ,
\end{equation}
where $T$ are elements of the undistorted pristine crystal's space group, i.e., translations, rotations, and other crystallographic symmetry operations.
We then define a graph adjacency matrix {\bf A}, where
\begin{equation}
A_{ij}=\left\{
\begin{array}{ll}
	1 \qquad \qquad i \neq j, D_{ij}<d_\textrm{max} \\
	0 \qquad \qquad \qquad \textrm{otherwise}
\end{array},
\right.
\end{equation}
with $d_\textrm{max}$ being a user specified distance used to determine whether any two muon sites are {\it connected}.  The matrix {\bf A} defines a graph where the muon sites represent nodes and if $A_{ij}=1$, there is an edge between nodes $i$ and $j$ whereas if $A_{ij}=0$ there is none.
Clusters of distinct sites correspond to the connected components \cite{Hopcroft:1973:AEA:362248.362272} of this graph, determined using the NetworkX library \cite{SciPyProceedings_11}.  A connected component of an undirected graph is a subgraph in which any two nodes are connected to each other by at least one path, and whose nodes are connected to no other nodes in the rest of the supergraph (see Figure~\ref{fig:fig3}).  As seen in Figure~\ref{fig:fig3}, it is not necessary for each node in a component to be directly connected to each other node in the same component.  This can be helpful in cases where the potential energy landscape for the muon has broad minima.  Depending on the force tolerance used in the calculations, the shallow gradient of the potential near such a minimum could result in a number of seemingly final muon positions around the true minimum, which are not all within $d_\mathrm{max}$ of each other, but are spread out more widely.  However, for two sites at the extremities of this single minimum it should be possible to form a path between them using other sites that relaxed towards the same minimum, , assuming a sufficiently dense sampling of candidate muon sites.  For sites belonging to distinct potential minima it should not be possible to form a connected path between them in this manner and hence the connected-component algorithm will correctly identify them as being distinct.

The distance $d_\mathrm{max}$ is specified by the user, with no \textit{a priori} optimal value, though in most cases $0.2 < d_\mathrm{max} < 1$~\AA~will produce sensible results.  In practice, the user will vary $d_\mathrm{max}$ until they get a satisfactory clustering.  If $d_\mathrm{max}$ is very large then all of the muon positions will be connected and the algorithm will return only a single cluster.  Conversely, a very small $d_\mathrm{max}$ will lead to a large number of sparsely populated clusters that will not only be unwieldy to carry forward for future analysis, but might not actually be physically meaningful, being, perhaps, merely an artefact of poor DFT force convergence around one or more shallow potential minima, as mentioned above.
Once the muon sites have been divided into sensible distinct clusters, the space group symmetry operations of the crystal can be used to bring the relaxed muon positions as close as possible to the position of the lowest energy muon site within the same cluster.  This results in a `clumping' of muon positions in space and can make it easier to visualise the distinct clusters found by the algorithm. 

\subsection{Correcting dipolar fields for distorted structures}\label{distorted}

\begin{figure}[t]
\centering
\includegraphics[width=\columnwidth]{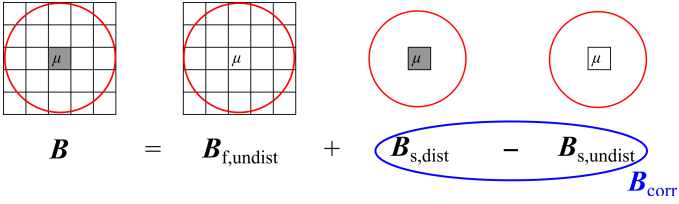}
\caption{Schematic indicating the unit cells included in the dipolar sums and the Lorentz spheres (red) for each of the calculations used to obtain the dipolar field corrected for muon induced distortions.  Shaded cells correspond to the relaxed, muon-distorted geometry obtained from muon site calculations whereas empty cells are undistorted, pristine-crystal cells.}
\label{fig:fig4}
\end{figure}

In most cases the atomic distortions induced by the muon are short-ranged and persist over only a few angstroms \cite{moeller}.  A good approximation to the magnetic moments seen by the muon can therefore be obtained by embedding a relaxed supercell containing the muon within a matrix of undistorted unit cells.  Provided the supercell used for structural relaxation was sufficiently large, any crystallographic distortions outside of the simulated supercell should be small enough to neglect.  Calculating the dipolar magnetic field at the muon site due to a moment distribution constructed in this manner is achieved by splitting the dipolar field into three components
\begin{equation}\label{fields}
\textit{\textbf{B}}=\textit{\textbf{B}}_\mathrm{f, undist}+\textit{\textbf{B}}_\mathrm{s, dist}-\textit{\textbf{B}}_\mathrm{s, undist},
\end{equation} 
where $\boldsymbol{B}_\mathrm{f, undist}$ is the full undistorted dipolar field calculated for a large number of unit cells (sufficiently large to ensure convergence) and is what is returned by carrying out a calculation assuming an undistorted crystallographic structure, i.e., before introducing the muon. $\boldsymbol{B}_\mathrm{s, dist}$ and $\boldsymbol{B}_\mathrm{s, undist}$ represent the local dipolar fields due to only the magnetic moments within the simulation supercell for the distorted and undistorted supercells, respectively.  Thus the difference between these two terms gives the correction to the total dipolar field,  $\boldsymbol{B}_\mathrm{corr}=\textit{\textbf{B}}_\mathrm{s, dist}-\textit{\textbf{B}}_\mathrm{s, undist}$, due to the fact that the ions in the simulation supercell are displaced by the presence of the muon. This procedure is illustrated by the schematic in Fig.~\ref{fig:fig4}.  Note that the Lorentz field is a property of only the magnetic moments outside the Lorentz sphere \cite{Kittel}, which is larger than the simulation supercell in all cases. This means that the Lorentz fields of the $\boldsymbol{B}$ and $\boldsymbol{B}_\mathrm{f,undist}$ contributions are the same, since we assume that magnetic moments outside the simulation supercell, and thus also outside the Lorentz sphere, are left unchanged. Furthermore, in the case of $\boldsymbol{B}_\mathrm{s,dist}$ and $\boldsymbol{B}_\mathrm{s,undist}$ contributions there is no Lorentz field due to outside moments, since the fields are calculated from all of the supercell magnetic moments.

For $\boldsymbol{B}_\mathrm{s,dist}$ and  $\boldsymbol{B}_\mathrm{s,undist}$ only the moments in a single supercell are included and thus the system is no longer periodic.  The supercell is recentred on the muon by translating individual magnetic ions by Bravais lattice vectors, such that the distances between the muon and each of the translated ions corresponds to the shortest possible distance within the periodic structure. This is also done for an undistorted supercell of the same size to allow a direct comparison within $\boldsymbol{B}_\mathrm{corr}$.  A  complication can arise if the displacements of the ions are such that they cross the boundary of this recentred cell, which results in an anomalously large change in their positions when compared to the undistorted cell.  To account for this possibility, each of the ions in the undistorted cell are translated by lattice vectors such that they are as close as possible to the corresponding ion in the distorted cell.  
This method is only implemented for commensurate magnetic structures, for which the magnetic unit cell can be fit inside the simulation cell. However, it can be extended to incommensurate structures in principle by considering the full $B$ from Eq.~\eqref{fields} corresponding to all of the magnetic moment configurations in the unit cell generated by the propagation vector.

\section{Usage}\label{usage}
In this section we demonstrate the usage of MuFinder, by considering the example of $\mu^+$ in the insulating fluoride CoF$_2$. The muon sites in this system were previously calculated in one of the earliest DFT+$\mu$ calculations \cite{moeller}, with fluorides being ideal systems to study in this context because the calculated site(s) can be verified experimentally by observing the characteristic F--$\mu$--F oscillations due to dipolar coupling leading to quantum entanglement between the muon and the $^{19}$F nuclear spins \cite{PhysRevB.33.7813}. We work through the four tabs of the GUI in order, following the workflow presented in Fig.~\ref{fig:fig1}.

\subsection{Generation}
The first step in a muon site calculation is to generate a set of initial muon positions that form the initial guesses for the muon sites; this is done in the Generation tab. For an unbiased search that doesn't rely on any prior knowledge of likely stopping sites, we will want to randomly sample positions within the unit cell, which is accomplished using the algorithm described in section \ref{gen_alg}. The results of doing this for CoF$_2$, with $r_\mathrm{muon}=0.5$~\AA{} and $r_\mathrm{atom}=1.0$~\AA, is shown in Fig.~\ref{fig:fig2}. Once the set of initial positions has been generated, these can be relocated to a symmetry-reduced subspace using a similar method as for relaxed muon positions, as described in section \ref{clustering}. 
As seen in Fig.~\ref{fig:fig5}, only a wedge comprising 1/16 of the volume of the unit cell needs to be considered in this case to sample all of the symmetry-distinct positions. Finally, we generate a set of CASTEP \texttt{.cell} files for use in structural relaxations, with each initial muon position corresponding to a different cell file. We use a $2 \times 2 \times 2$ supercell of CoF$_2$ for these calculations, as the conventional unit cell is not large enough to ensure that the muon is isolated from its periodic images in DFT calculations.

\subsection{Run}
The next step is to carry out a geometry optimisation calculation for each of the initial muon positions, which can be done via the Run tab. For our example of CoF$_2$ we ran these calculations remotely on a high performance computing (HPC) cluster by providing MuFinder with the appropriate submission script template, but it is also possible to run these calculations locally, which may be more suitable for smaller systems or those containing fewer electrons.  Various parameters that are to be used for these calculations can be specified through the GUI; here we treat the system as spin-polarised and use the PBE exchange-correlation functional \cite{PhysRevLett.77.3865} and the ensemble density functional theory (EDFT) solver. All other parameters are left to take their default values. Remote calculations can be submitted, monitored and managed through the GUI, and once a job has finished successfully the output files required for subsequent analysis are copied to the local machine. MuFinder will flag jobs that exit the queue without writing the expected output files so that the user can investigate these. Jobs that end prematurely due to hitting the wall time limit can be resubmitted through the GUI as a continuation.

\subsection{Analysis}
\begin{figure}[ht]
\centering
\includegraphics[width=0.9\columnwidth]{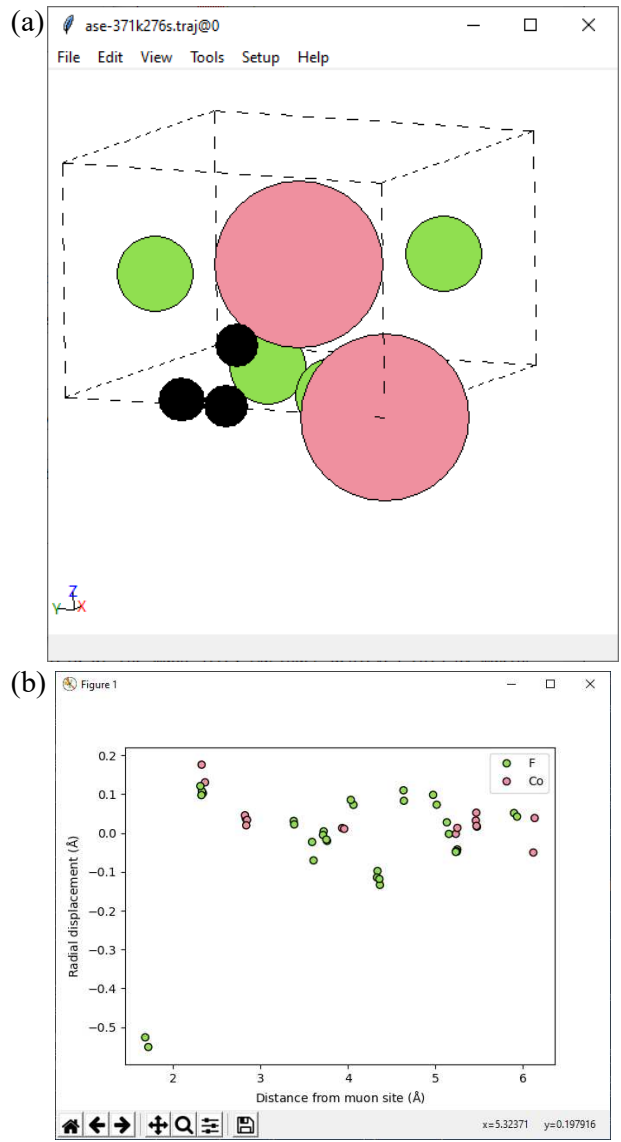}
\caption{(a) Clusters of muon sites (represented by black spheres) in CoF$_2$, shown within the conventional unit cell. (b) Radial displacements of ions as a function of their distances from the lowest energy muon site.}
\label{fig:fig5}
\end{figure}

Once all of the structural relaxations have finished, we can perform clustering operations on the set of muon sites obtained in the Analysis tab. Here we use the clustering methodology based on graph theory, which clusters sites based on their positions within the unit cell by taking symmetry into account as described in detail in section \ref{clustering}. MuFinder can also cluster muon sites via $k$-means clustering as implemented in the Soprano code and described in Ref.~\cite{liborio}, which also takes the energies of the relaxed structures into account in addition to the muon position. Once the clusters have been identified, symmetry operations are used to relocate muons such that muons within the same cluster are brought closer together in physical space (`clumped'), allowing a better visualisation of the distinct stopping sites, as also described in section \ref{clustering}. The clusters of muon sites obtained for CoF$_2$ with $d_\mathrm{max}=0.25$~\AA{} are shown in Fig.~\ref{fig:fig5}(a). We identify three distinct clusters, with the lowest energy site involving the muon sitting at the position with fractional coordinates (0, 0.5, 0) (relative to the conventional unit cell), in agreement with the site previously obtained both from experiment \cite{PhysRevB.30.186} and from DFT \cite{moeller}. The other candidate muon sites are 0.34 and 0.82 eV higher in energy than this site and are therefore unlikely to be occupied, and are also incompatible with experiment.

Further analysis of individual muon sites is also possible within the Analysis tab. By comparing the pristine and relaxed structures, MuFinder can determine the muon-induced displacement of the host atoms as a function of distance from the muon site. The muon-induced distortions for the lowest-energy site in CoF$_2$ are shown in Fig.~\ref{fig:fig5}(b). We see that the muon attracts the neighbouring F atoms towards itself and these are displaced by 0.50 and 0.57 \AA, respectively. The nearby Co$^{2+}$ ions are displaced by 0.14 and 0.17 \AA{} respectively away from the muon and this has consequences for the dipolar magnetic field seen by the muon, as shown in section \ref{dipole}. The displacements can be seen to decay as a function of distance away from the muon site, as expected. Examining the magnitude of the displacements at the furthest distances from the muon site can be helpful for evaluating whether a sufficiently large supercell has been used, as large displacements at these points indicate that the muon could be sensitive to the displacement field induced by its implicit periodic images in DFT calculations, and that a larger supercell should thus be used to avoid this finite-size artefact.

\subsection{Dipole Field}\label{dipole}
The Dipole Field tab allows us to make contact with the internal magnetic fields measured by the muon in experiment by calculating the dipolar magnetic field at the muon site. MuFinder achieves this by making use of the MuESR Python library \cite{muesr}. In the magnetically ordered state of CoF$_2$, the Co magnetic moments lie along the $c$ axis and are ordered antiferromagnetically within the unit cell, with magnitudes of 2.64 $\mu_\mathrm{B}$ \cite{PhysRev.137.A982}. The calculated dipolar field due to these ions at the lowest energy muon site is shown in Table \ref{CoF2_fields}, where it is compared with experiment \cite{PhysRevB.30.186} and the calculations of M\"{o}ller {\it et al.} \cite{moeller}. A benefit of the structural relaxation method is that, in addition to the muon site, it reveals changes in the positions of nearby atoms in the host resulting from the presence of the muon. MuFinder takes advantage of this by including the effects of distortions to the magnetic ions when evaluating the field at the muon site, using the approach described in section \ref{distorted}. For the present example, taking these distortions into account produces dipolar fields that are significantly smaller than those obtained from an undistorted structure, with the field obtained when incorporating these distortions nearly identical to the one reported by M\"{o}ller {\it et al.} \cite{moeller}. These fields are slightly smaller than that obtained from experiment, which lies in between those calculated for distorted and undistorted structures.

\begin{table}[b]
\centering
\caption{\label{CoF2_fields}Dipolar field at the $\mu^+$ stopping site calculated by M\"{o}ller {\it et al.} \cite{moeller} and the field obtained using the MuFinder program, along with the local magnetic field measured experimentally \cite{PhysRevB.30.186}.}
\begin{tabular}{l c c}
	\hline 
	& \multicolumn{2}{c}{$B$ (T)}  \\
	\hline
	Experiment \cite{PhysRevB.30.186} & \multicolumn{2}{c}{0.228} \\
	\hline
	& undistorted & distorted \\
	\hline
	M\"{o}ller {\it et al.} \cite{moeller} &	0.265	&	0.208	\\
	MuFinder &	0.265&	0.207	\\
	\hline
\end{tabular}
\end{table}

\section{Conclusion}\label{conclusion}

MuFinder allows muon site calculations using the DFT structural relaxation method to be carried out through a user-friendly GUI. The user can generate initial muon position and can then relax these starting structures using CASTEP, with the program providing an interface to carry out these calculation either locally or on a remote machine or cluster. Once calculated, the resulting muon sites can be clustered using a symmetry-aware algorithm and then further analysed, such as by evaluating the local magnetic field expected at the muon site for candidate magnetic structures. The entire procedure, starting from the crystal structure and ending with a set of candidate muon sites and predictions for the local fields can be done within the GUI, making it highly accessible to non-experts in electronic structure methods.

At its core, MuFinder represents the automation of a workflow practised by several research groups carrying out DFT+$\mu$ calculations. It makes use of previously developed tools such as MuESR, with the fact that these are also written in Python allowing these to be seamlessly integrating into the workflow. Much of the heavy lifting is done using ASE, which allows the muonated structures to be passed around between different parts of the program. This will also facilitate the integration of further analysis tools as the program continues to be developed, as these will likely require the relaxed geometry of the system plus implanted muon as an input.

MuFinder has already been used successfully to determine muon stopping sites in a variety of systems, in each case significantly enhancing the interpretation of the experimental $\mu^+$SR results. For the skyrmion-hosting systems Cu$_2$OSeO$_3$ and Co$_x$Zn$_y$Mn$_{20-x-y}$, the calculated muon sites were used to obtain the distributions of internal field expected for the skyrmion lattice and surrounding phases \cite{PhysRevB.103.024428}. In the kagom\'{e} antiferromagnet barlowite, DFT calculations of the muon site enabled the interpretation of the $\mu^+$SR spectra in terms of the formation of both $\mu^+$--F and $\mu^+$--OH complexes \cite{Tustain2020}. For the transition metal dichalcogenide 1T-TaS$_2$, the muon sites, calculated using MuFinder, provided insight into the sensitivity of the muon to the magnetic states of adjacent layers \cite{Manas-Valero2021}. 
These techniques have also been successfully applied to molecular magnets, such as the staggered spin chain material [pym-Cu(NO$_3$)$_2$(H$_2$O)$_2$] \cite{PhysRevB.103.L060405}, with the complicated structures of these systems, containing many atoms, often leading to large numbers of candidate muon sites. Muon site calculations on several superconductors \cite{huddart2021intrinsic} were able to rule out the possibility of the time-reversal symmetry breaking detected in $\mu^+$SR being an artifact of muon-induced distortions, which had been a longstanding concern in this field. These examples highlight the kinds of analyses that one can hope to achieve after obtaining muon stopping sites using a tool such as MuFinder.

Despite the structural relaxation method being a fairly well established approach for determining muon sites, research is ongoing to determine cases were this approach is likely to be insufficient, such as when quantum effects are present. While such methods remain in their infancy it is hoped that successful approaches could be integrated into MuFinder as they are developed. Another aspect that is not currently addressed by MuFinder is the contribution of the hyperfine field at the muon site to the effective local magnetic field seen by the muon. The hyperfine field is difficult to compute in general, though we note progress has recently been made in addressing this for select materials \cite{PhysRevB.97.174414,PhysRevMaterials.5.044411}.  MuFinder has the potential to act as a platform for bringing these advances in computational science into the reach of the experimentalist, thereby establishing muon site calculations as a routine part of conducting a $\mu^+$SR experiment. 

MuFinder is distributed in the form of binaries for either Windows or Ubuntu, with the Python source code also freely available under GNU GPLv3. The program, along with a complete manual, can be accessed via Ref.~\cite{gitlab}.

\section*{Acknowledgments}
This work is supported by EPSRC (UK), under
Grants No. EP/N024028/1, No. EP/N024486/1 and No. EP/N032128/1. M. G. acknowledges the support of the Slovenian Research Agency under Projects No. Z1-1852 and J1-2461. We acknowledge
computing resources provided by STFC Scientific
Computing Department's SCARF cluster and Durham
Hamilton HPC. B.\,M.\,H and A.\,H.\,M. thank STFC for support via studentships. We thank S.J. Blundell, R. De Renzi, L. Liborio and S. Sturniolo for useful discussions.





\bibliographystyle{elsarticle-num}
\bibliography{references}







\end{document}